\documentclass[aps,prb,showpacs,twocolumn,superscriptaddress,floatfix]{revtex4}

\usepackage{graphicx}
\usepackage{dcolumn}
\usepackage{bm}
\usepackage{amsmath}
\usepackage{amssymb}
\usepackage{color}
\usepackage{longtable}
\usepackage{here}

\begin{document}

\renewcommand{\vec}[1]{\mbox{\boldmath $#1$}}


\title{Microscopic examinations of Co valences and spin states in electron-doped LaCoO$_3$}


\author{Keisuke Tomiyasu}
\email[Electronic address: ]{tomiyasu@m.tohoku.ac.jp}
\affiliation{Department of Physics, Tohoku University, Aoba, Sendai 980-8578, Japan}
\author{Syun-Ichi Koyama}
\affiliation{Department of Physics, Tohoku University, Aoba, Sendai 980-8578, Japan}
\author{Masanori Watahiki}
\affiliation{Department of Physics, Tohoku University, Aoba, Sendai 980-8578, Japan}
\author{Mika Sato}
\affiliation{Department of Physics, Tohoku University, Aoba, Sendai 980-8578, Japan}
\author{Kazuki Nishihara}
\affiliation{Department of Physics, Tohoku University, Aoba, Sendai 980-8578, Japan}
\author{Yuki Takahashi}
\affiliation{Department of Physics, Tohoku University, Aoba, Sendai 980-8578, Japan}
\author{Mitsugi Onodera}
\affiliation{Department of Physics, Tohoku University, Aoba, Sendai 980-8578, Japan}
\author{Kazuaki Iwasa}
\affiliation{Department of Physics, Tohoku University, Aoba, Sendai 980-8578, Japan}
\author{Tsutomu Nojima}
\affiliation{Institute for Materials Research, Tohoku University, Aoba, Sendai 980-8577, Japan}
\author{Hiroyuki Nojiri}
\affiliation{Institute for Materials Research, Tohoku University, Aoba, Sendai 980-8577, Japan}
\author{Jun Okamoto}
\affiliation{National Synchrotron Radiation Research Center, Hsinchu 30076, Taiwan}
\author{Di-Jing Huang}
\affiliation{National Synchrotron Radiation Research Center, Hsinchu 30076, Taiwan}
\author{Yuuichi Yamasaki}
\affiliation{Condensed Matter Research Center and Photon Factory, Institute of Materials Structure Science, High Energy Accelerator Research Organization, Tsukuba, Ibaraki 305-0801, Japan}
\author{Hironori Nakao}
\affiliation{Condensed Matter Research Center and Photon Factory, Institute of Materials Structure Science, High Energy Accelerator Research Organization, Tsukuba, Ibaraki 305-0801, Japan}
\affiliation{CREST, Japan Science and Technology Agency (JST), Tokyo, 102-0076 Japan}
\author{Youichi Murakami}
\affiliation{Condensed Matter Research Center and Photon Factory, Institute of Materials Structure Science, High Energy Accelerator Research Organization, Tsukuba, Ibaraki 305-0801, Japan}


\date{\today}

\begin{abstract}
We studied the Co valences and spin states in electron-doped LaCo$_{1-y}$Te$_{y}$O$_3$ by measuring x-ray absorption spectra and electron spin resonance. The low-temperature insulating state involves the low-spin Co$^{3+}$ ($S=0$) and the high-spin Co$^{2+}$ state, which is described by $g=3.8$ and $j_{\rm eff}=1/2$. The results, in concurrence with the electron--hole asymmetry confirmed in electrical resistivity, coincide with a spin-blockade phenomenon in this system. Further, we discuss the $g$ factor in terms of the strong covalent-bonding nature and consider multiple origins of this phenomenon.
\end{abstract}


\maketitle

\section{Introduction}
%

\begin{figure}[b]
\begin{center}
\includegraphics[width=0.88\linewidth, keepaspectratio]{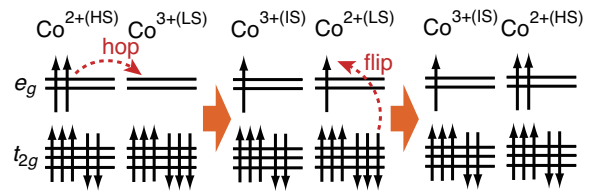}
\end{center}
\caption{\label{fig:scheme} (Color online)
Scheme of the multistep processes of electron hopping from Co$^{2+{\rm (HS)}}$ to Co$^{3+{\rm (LS)}}$. If Co$^{3+{\rm (IS)}}$ is also unfavorable, the spin-state change (flip) to HS or LS is added for Co$^{3+}$, which further suppresses the total hopping probability. }
\end{figure}

The many-body problem is ubiquitous in a variety of physics fields. In condensed matter physics, correlated electron systems have been shown to exhibit potentially useful phenomena, such as spin-ordered Mott insulation, charge-ordered insulation, magnetoresistance, and superconductivity,~\cite{1972_Anderson} via spin, charge, orbital, and lattice. 
Recently, remarkable spin-dependent charge transfer called spin blockade was reported.
One type of spin blockade is observed in quantum-dot systems, in which the Pauli exclusion principle forbids electron hopping between two dots with the same spin (Pauli blockade).~\cite{2002_Ono}
Another type originates from a prominent concept, the utilization of unique spin-state variability/invariability, as explained below, and is considered to be active in some perovskite-type cobaltites.~\cite{2004_Maignan, 2005_Taskin, 2009_Chang} The fivefold 3$d$ orbitals split to $t_{2g}$ and $e_g$ manifolds in CoO$_6$ octahedra. While Co$^{3+}$ ($d^6$) may have low-spin (LS, $S=0$), high-spin (HS, $S=2$), or intermediate-spin (IS, $S=1$) states as in LaCoO$_3$,~\cite{1996_Korotin} Co$^{2+}$ ($d^7$) is semi-empirically known to prefer an HS state ($S=3/2$).~\cite{2004_Maignan, 1970_Sugano} Based on this Co character, the electron hopping from Co$^{2+{\rm (HS)}}$ to Co$^{3+{\rm (LS)}}$ leads to the unfavorable Co$^{2+{\rm (LS)}}$ ($S=1/2$) state accompanied by a simultaneous second process in which the spin state changes to Co$^{2+{\rm (HS)}}$, as shown in Fig.~\ref{fig:scheme}; therefore, the total hopping probability will be considerably suppressed.~\cite{2004_Maignan}
Hereinafter, {\it regardless of its origin}, we generally refer to the suppression phenomena of charge transfer between the sites with different spin states as spin-state blockade.

%
%

Spin-state blockade was experimentally observed in CoO$_2$ conduction planes. For the ordered oxygen-deficient perovskite semiconductor GdBaCo$_2$O$_{5.5+\delta}$ with Co$^{3+{\rm (LS)}}$O$_6$ octahedra and Co$^{3+{\rm (IS)}}$O$_5$ pyramids, hole doping (Co$^{4+}$; $\delta>0$) greatly decreases the electrical resistivity, whereas electron doping (Co$^{2+}$; $\delta<0$) does not.~\cite{2005_Taskin} This electron--hole asymmetry confirms the spin-state blockade of electron doping.~\cite{2005_Taskin} Further, in the double-perovskite insulator La$_{1.5}$Sr$_{0.5}$CoO$_4$, the alternately charge-ordered Co$^{2+}$ and Co$^{3+}$ (nominally Co$^{2.5+}$) have spin states revealed to be HS and LS, respectively, by using soft x-ray absorption spectroscopy, demonstrating that the condition of spin-state blockade is satisfied.~\cite{2009_Chang} Thus, material designs explicitly utilizing the spin-state blockade could be explored.~\cite{2004_Maignan, 2005_Taskin, 2009_Chang}

Then, one of the intriguing issues is to examine the spin-state blockade phenomena in an electron-doped system with {\it sensitive} Co spin-state variability; the best candidate for the matrix is perovskite-type LaCoO$_3$ in its lowest-temperature (LT) range, as mentioned in the conclusion of Ref.~[\cite{2004_Maignan}].
This material consists of a three-dimensional network of corner-sharing CoO$_6$ octahedra. At room temperature (RT), this system is electrically semiconductive and is magnetically composed of thermally excited Co$^{3+{\rm (HS)}}$ and possibly Co$^{3+{\rm (IS)}}$ with a ground state of Co$^{3+{\rm (LS)}}$. As the temperature decreases, spin crossover (spin-state transition) occurs from approximately 100 K to 30 K, and the system finally enters the highly insulating LT range with only the nonmagnetic ground state of Co$^{3+{\rm (LS)}}$.~\cite{1964_Heikes, 1972_Bhide, 1996_Korotin, 2006_Haverkort}
Such a spin-state transition is sensitively induced not only by temperature but also by chemical substitution, magnetic field, and pressure.~\cite{2009_Sato, 1997_Asai, 1996_Yamaguchi}

\begin{figure}[htbp]
\begin{center}
\includegraphics[width=0.77\linewidth, keepaspectratio]{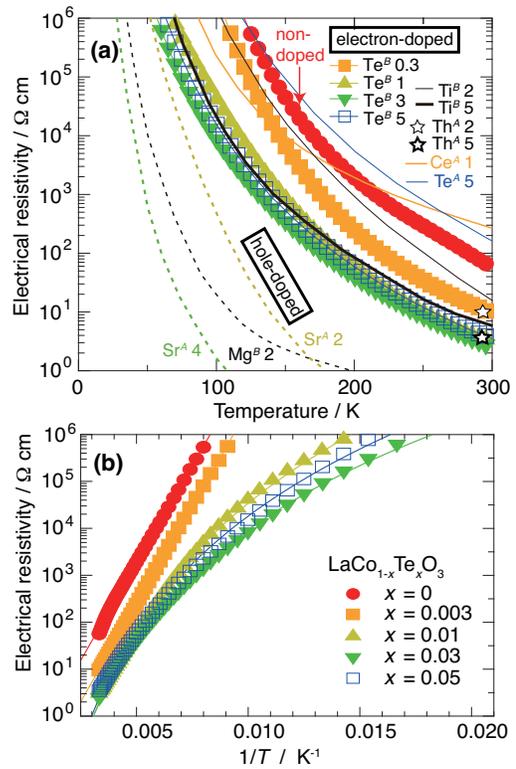}
\end{center}
\caption{\label{fig:rhoT} (Color online)
(a) Temperature dependence of the electrical resistivity in pure and substituted LaCoO$_3$ systems. The symbols with lines show our measured data for LaCo$_{1-y}$Te$_{y}$O$_3$. The other symbols and lines show the reference data,~\cite{1962_Gerthsen, 2004_Maignan, 2004_Kriener, 2008_Zheng, 2008_Jirak} which are normalized so that the pure LaCoO$_3$ data coincide with the $y=0$ data at each temperature. The superscripts $A$ and $B$ denote the La-site and Co-site substitutions, respectively. The numbers written after elements denote the substitution concentration in percentage.
(b) Electrical resistivity versus inverse temperature in LaCo$_{1-y}$Te$_{y}$O$_3$. The straight lines denote the Arrhenius thermal-activation type and the curves denote the power-law type: $\rho \propto 1/T^{\nu}$ ($\nu = 8$ to 9).}
\end{figure}

Thus far, spin-state blockade activation has been discussed for various electron-doping elemental substitutions in LaCoO$_3$. In fact, the electron--hole asymmetry in electrical resistivity, the first necessary condition of spin-state blockade, has been reported, for example, in comparisons between electron-doped Th, Ce, and Te substitutions and hole-doped Sr substitution at the La site~\cite{1962_Gerthsen, 2004_Maignan, 2008_Zheng} and comparisons between electron-doped Ti substitutions and hole-doped Mg substitutions at the Co site~\cite{2008_Jirak}. As shown in Fig.~\ref{fig:rhoT}(a), overall, electron doping does not decrease the electrical resistivity, in contrast to hole doping.

However, the Co$^{3+({\rm LS})}$--Co$^{2+({\rm HS})}$ coexistence, the second necessary condition of spin-state blockade, remains experimentally uncertain;
the spin states of Co$^{3+}$ and Co$^{2+}$ have been discussed on the basis of magnetization measurements, which are somewhat confusing and controversial.
For example, while Co$^{3+}$ is interpreted as an LS state,~\cite{2008_Zheng} a considerable amount of IS and/or HS states are expected to exist.~\cite{2008_Alvarez-Serrano, 2008_Jirak, 2008_Hejtmanek} Further, although the very high stability of Co$^{2+ {\rm (HS)}}$ is often assumed,~\cite{2004_Maignan, 2005_Taskin, 2008_Zheng, 2009_Chang} it has been suggested that Co$^{2+({\rm LS})}$ can be realized.~\cite{1999_Nakatsugawa, 2008_Hejtmanek, 2008_Alvarez-Serrano} In addition, there might be charge homogeneity (itinerancy) rather than the Co$^{2+}$--Co$^{3+}$ good charge separation.~\cite{2008_Zheng, 2013_Augustinsky}
Thus, not only magnetization measurements but also microscopic experiments such as XAS are necessary to determine the Co valences and spin states.~\cite{2009_Chang} In fact, the understanding of the Co states is quite difficult because of the complex unquenched-orbital angular momenta and strong covalent-bonding nature, in addition to the valence and spin-state degree of freedom.~\cite{1957_Kanamori_a, 1996_Korotin, 2002_Noguchi, 2006_Podlesnyak} 

%
%

Suitable electron-doped LaCoO$_3$ systems are rather limited. A variety of hole-doped La$_{1-x}$$A_{x}$CoO$_3$ systems have been obtained ($A=$ Ca, Sr, Ba; $0 \leq x \leq 1$).~\cite{2004_Kriener, 2003_Wu, 2011_Long} Although the $A=$ Ce, Th, and Te substitutions are expected to dope an electron with $A^{4+}$, single-phase Ce is relatively difficult to obtain in bulk, and the radioactivity of Th is difficult to handle.~\cite{1989_Tabata, 2005_Fuchs, 1987_Tabata, 2008_Zheng} Moreover, the Co, Ce, Th, and Te valences have not been evaluated using methods such as x-ray absorption spectroscopy (XAS) to evaluate the electron doping. In LaCo$_{1-y}$$M_{y}$O$_3$ systems ($M$ = Ti; $0 \leq y \leq 0.5$),~\cite{1999_Nakatsugawa, 2005_Cairns, 2008_Jirak, 2008_Hejtmanek, 2008_Alvarez-Serrano, 2010_Robert} XAS spectra showed that the average Co valence decreases from $3+$ to $2.5+$ ($y=0.5$)~\cite{2010_Robert} with electron doping using nominal Ti$^{3.5+}$ ($d^{0.5}$; magnetic) substitution.

An electron-doped system, LaCo$_{1-y}$Te$^{6+}_{y}$O$_3$ (Te$^{6+}$: $d^{10}$; nonmagnetic), is relatively easy to synthesize in a solid-state reaction method and allows us to avoid the complexity with magnetic $M$ atoms. Further, the ionic radius of Te$^{6+}$ has been reported to be between those of Co$^{3+{\rm (LS)}}$ and Co$^{3+{\rm (HS)}}$ ($r_{{\rm Te}^{6+}}=0.56$ {\AA}, $r_{{\rm Co}^{3+{\rm (LS)}}}=0.545$ {\AA}, $r_{{\rm Co}^{3+{\rm (HS)}}}=0.61$ {\AA}),~\cite{1976_Shannon} suggesting size matching for Co-site substitution to minimize the difficulties accompanying local structural disorder.~\cite{2007_Pinta} The electrical resistivity ($\rho$) [symbols with lines in Fig.~\ref{fig:rhoT}(a)] is also consistent with the tendency of electron--hole asymmetry.
On the other hand, the $6+$ valence is expected to give rise to high Coulomb scattering on $\rho$. Thus, we checked the total disorder effect using the $\rho$ versus $1/T$ analysis [Fig.~\ref{fig:rhoT}(b)], as performed for LaCo$_{1-x}$Ti$_{x}$O$_3$.~\cite{2008_Jirak} In the Ti system, as $x$ increases, $\rho$ changes from the simple Arrhenius thermal-activation type to a characteristic power-law type -- $\rho \propto 1/T^{\nu}$ ($\nu = 8$ to 10) -- which is considered to reflect the local disorder effect.~\cite{2008_Jirak} For LaCo$_{1-y}$Te$_{y}$O$_3$, an identical power-law behavior is obtained, as shown in Fig.~\ref{fig:rhoT}(b). Hence, the total disorder effect on $\rho$ in the Te system will be comparable to that in the Ti system. 

In this study, we microscopically show the Co$^{3+({\rm LS})}$--Co$^{2+({\rm HS})}$ coexistence in LaCo$_{1-y}$Te$_{y}$O$_3$ by measuring XAS spectra and electron spin resonance (ESR). First, the Te and Co valences are estimated using XAS at the Te $L_{1}$ and Co $K$ edges in the hard x-ray region. Second, the Co valences and spin states are further revealed using XAS at the Co $L_{3,2}$ and O $K$ edges in the soft x-ray region. Third, the Co$^{2+}$ spin state is determined with the $g$ factor by using ESR. Last, we discuss several origins of the spin-state blockade.

%
\section{Experiments}
\begin{figure}[htbp]
\begin{center}
\includegraphics[width=0.95\linewidth, keepaspectratio]{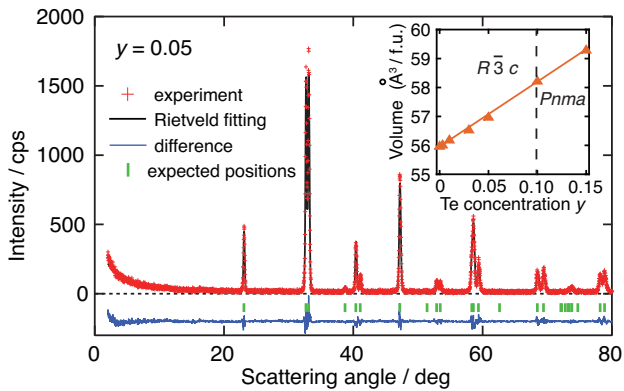}
\end{center}
\caption{\label{fig:xrd} (Color online) Typical x-ray diffraction data. The inset shows the Te concentration ($y$) dependence of lattice volume. }
\end{figure}

%
LaCo$_{1-y}$Te$_{y}$O$_3$ samples were synthesized by using a stardanrd solid-state reaction method. The starting materials, La$_2$O$_3$ (99.99{\%} purity), Co$_3$O$_4$ (99.9{\%}), and TeO$_2$ (99.9{\%}) powders, were purchased from Kojundo Chemical Laboratory Co.Ltd. Stoichiometric mixtures of them were ground throughly, pelletized, and heated on ZrO$_2$ plates at 1200$^{\circ}$C in air for 12 h twice (24 h in total). 
No Bragg reflections, other than those of the perovskite structure, were detected from all the samples in the Cu-K$\alpha$ x-ray diffraction patterns measured at room temperature. The typical data and Rietveld fitting were shown in Fig.~\ref{fig:xrd}. The Rietveld fitting was performed using Z-Rietveld software.~\cite{Z_2009, Z_2012} The obtained lattice volume is almost the same as the representative reported value at $y=0$~\cite{2002_Radaelli} and monotonically increases with increasing $y$, as shown in the inset in Fig.~\ref{fig:xrd}. The composition ratio LaCo$_{1-y}$Te$_{y}$O$_3$ is confirmed together with valences by XAS in the hard x-ray region in the next section [Figs.~\ref{fig:XAS}(a)--\ref{fig:XAS}(c)].
%
%

We remark the La$_{1-x}$Te$_x$CoO$_3$ samples previously reported.~\cite{2008_Zheng} In the non-substituted LaCoO$_3$, the lattice constants in the previous sample are much larger than the representative values; the former set at room temperature corresponds to the latter one above 600 K.~\cite{2008_Zheng, 2002_Radaelli} The reason might be related to the usage of Co$_2$O$_3$ and the rather low-temperature synthesis at 1050$^{\circ}$C,~\cite{2008_Zheng} which is, however, beyond the scope of this study. 

XAS measurements in the hard x-ray region were performed on the BL-3A beamline at the Photon Factory (PF) at KEK in Japan. The data were recorded in the bulk-sensitive fluorescence yield mode with fixed final energies at the Te $L_{1}$ and Co $K$-absorption edges. For the reference samples, polycrystalline TeO$_2$ (Te$^{4+}$), Te(OH)$_6$ (Te$^{6+}$), La$_2$CoO$_4$ (Co$^{2+}$), La$_{1.5}$Ba$_{0.5}$CoO$_4$ (Co$^{2.5+}$), and LaSrCoO$_4$ (Co$^{3+}$) were used.
XAS measurements in the soft x-ray region were performed on the BL-08B beamline at the National Synchrotron Radiation Research Center (NSRRC) in Taiwan and the BL-19B beamline at PF. Samples were cleaned with diamond files just before inserting them into a vacuum chamber with a base pressure of $5\times10^{-8}$ Torr. The data were also collected in the bulk-sensitive fluorescence yield mode. The surface-sensitive total electron yield data are not used in this study but were consistent with the fluorescence data. CoO was measured for the references of relative energy calibration and Co$^{2+({\rm HS})}$. The energy resolutions were approximately 0.3 eV and 0.2 eV for the Co $L_{3,2}$ edges and O $K$ edge, respectively.
ESR measurements were performed in pulsed magnetic fields at the Institute for Materials Research, Tohoku University. The frequencies are fixed in the range from 190 to 450 GHz. 
Direct-current magnetization was measured using superconducting quantum interference device (SQUID) magnetometers at the Center for Low Temperature Science, Tohoku University. Direct-current electrical resistivity was measured using the four-probe method.

\section{Results and discussion}
%

%
\subsection{XAS}
\begin{figure*}[htbp]
\begin{center}
\includegraphics[width=0.8\linewidth, keepaspectratio]{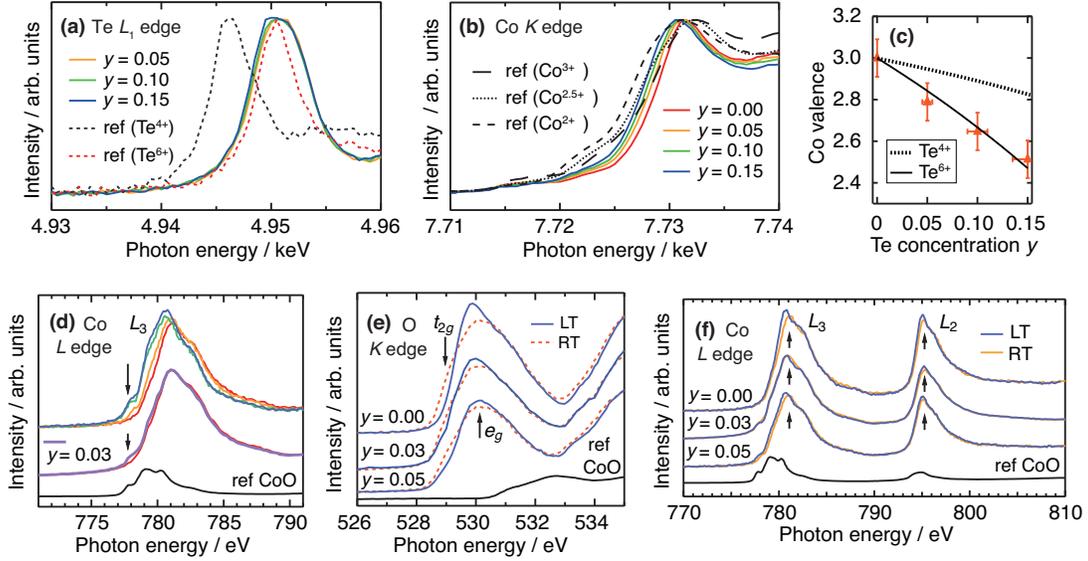}
\end{center}
\caption{\label{fig:XAS} (Color online)
(a,b) Te $L_1$-edge and Co $K$-edge XAS spectra measured at RT, respectively.
(c) Te-concentration dependence of the average Co valence. The positions of symbols are estimated from the Co $K$-edge energy shifting in (b). The solid line corresponds to the composition formula LaCo$^{v+}_{1-y}$Te$^{6+}_{y}$O$_{3}$, in which one Te dopes three electrons with $v=(3-6y)/(1-y)$ according to the charge-neutrality law. For comparison, the dotted line obtained for the composition formula LaCo$^{v+}_{1-y}$Te$^{4+}_{y}$O$_{3}$, $v=(3-4y)/(1-y)$, is shown.
(d) Co $L_{3}$-edge XAS spectra measured at RT. In the upper set of solid lines, the colors correspond to the same $y$ samples as those shown in (b). The arrows indicate the shoulders growing with increasing $y$.
(e,f) O $K$-edge and Co $L_{3,2}$-edge XAS spectra measured at RT and LT (15 K). The arrows indicate the difference from RT to LT.
In (d--f), the $y=0.03$ lines show the PF data and the other lines show the NSRRC data. }
\end{figure*}
%

Figures~\ref{fig:XAS}(a) and \ref{fig:XAS}(b) show XAS spectra of the LaCo$_{1-y}$Te$_{y}$O$_3$ samples and the reference samples in the hard x-ray region.
In Fig.~\ref{fig:XAS}(a), the Te $L_{1}$-edge spectra of LaCo$_{1-y}$Te$_{y}$O$_3$ are nearly coincident. Further, the peak positions of the LaCo$_{1-y}$Te$_{y}$O$_3$ samples are at much higher energies than that of the Te$^{4+}$ reference and are nearly equal to that of the Te$^{6+}$ reference. Thus, the Te valence of the samples is estimated to be nearly $6+$. 
This result is consistent with the fact that a number of perovskite $AB$O$_3$ materials with $B$-site Te$^{6+}$ substitution were reported.~\cite{1975_Politova} Meanwhile, the Te valence was previously assumed to be $4+$ in the $A$-site Te substitution system La$_{1-x}$Te$_{x}$CoO$_3$ as well as in Mn oxides,~\cite{2008_Zheng, 2003_Tan} which will be therefore attributable to the difference of substitution site. 
In Fig.~\ref{fig:XAS}(b), the Co $K$-edge spectra of LaCo$_{1-y}$Te$_{y}$O$_3$ shift to the lower-photon-energy side with increasing $y$ and those of references shift similarly on decreasing the average Co valence ($v$) from ${3+}$, confirming that $v$ proportionally decreases (electron doping) with Te substitution.
Further, the shifts are estimated to be $-1.2$ eV from $y=0$ to 0.15 and $-2.5$ eV from the reference Co$^{3+}$ to Co$^{2+}$. The relation of $v$ versus $y$, obtained from the shifting degree, is compared to those calculated from the composition formulae LaCo$^{v+}_{1-y}$Te$^{6+}_{y}$O$_3$ and LaCo$^{v+}_{1-y}$Te$^{4+}_{y}$O$_3$ in Fig.~\ref{fig:XAS}(c). The experimental relation is in agreement with that for LaCo$^{v+}_{1-y}$Te$^{6+}_{y}$O$_3$, indicating that electron-doped samples described by this composition formula were obtained.


Next, the Co states can be studied using XAS spectra in the soft x-ray region. Figure~\ref{fig:XAS}(d) shows the Co $L_3$-edge XAS spectra. The spectral lines consist of the $y=0$ base lines (matrix Co$^{3+}$) and the shoulders in the lower-photon-energy side increasing with $y$, as shown by the arrows. The shoulders are identical to the reference CoO line, which is consistent with the existence of Co$^{2+({\rm HS})}$.~\cite{2009_Chang}
Figure~\ref{fig:XAS}(e) shows the comparisons of O $K$-edge XAS spectra at RT and LT. This peak is separated from the CoO line in turn, and the lower- and higher-photon-energy parts correspond to the densities of unoccupied Co$^{3+}$ $t_{2g}$ and $e_g$ states (holes), respectively.~\cite{2012_Hu} In all the $y$ samples, the $t_{2g}$ holes decrease and the $e_{g}$ holes increase from RT to LT, indicating the Co$^{3+{\rm (HS)}}$-to-Co$^{3+{\rm (LS)}}$ spin-state transition. Furthermore, this transition corresponds to a slight sharpening of Co $L_{3,2}$-edge main peak tops from RT to LT,~\cite{2006_Haverkort} which is also observed [Fig.~\ref{fig:XAS}(f)].
Thus, these soft x-ray data [Figs.~\ref{fig:XAS}(d--f)] comprehensively indicate that the Co$^{2+{\rm (HS)}}$--Co$^{3+{\rm (LS)}}$ coexistence is realized towards LT.

%
\subsection{ESR}

To distinguish Co$^{2+{\rm (HS)}}$ and Co$^{2+{\rm (LS)}}$ more clearly, we estimated the $g$ factor by measuring ESR. The powder-averaged $g$ value is both experimentally and theoretically known to be approximately 4.3 for Co$^{2+{\rm (HS)}}$ ($j_{\rm eff} = 1/2$) and approximately 2 for Co$^{2+{\rm (LS)}}$ ($S = 1/2$);~\cite{1958_Low, 1963_Lines, 1994_Dance, 2013_Hoffmann} the Co$^{2+}$ spin state can be determined with this difference. Further, the aforementioned XAS measurements show that Co$^{3+}$ forms the LS state ($S=0$; nonmagnetic singlet), which exhibits no ESR signal at the lowest temperature below 20 K without the aid of thermal activation;~\cite{2002_Noguchi} hence, only Co$^{2+}$ signals are expected to be observed in the lowest-temperature range. 

\begin{figure*}[htbp]
\begin{center}
\includegraphics[width=0.95\linewidth, keepaspectratio]{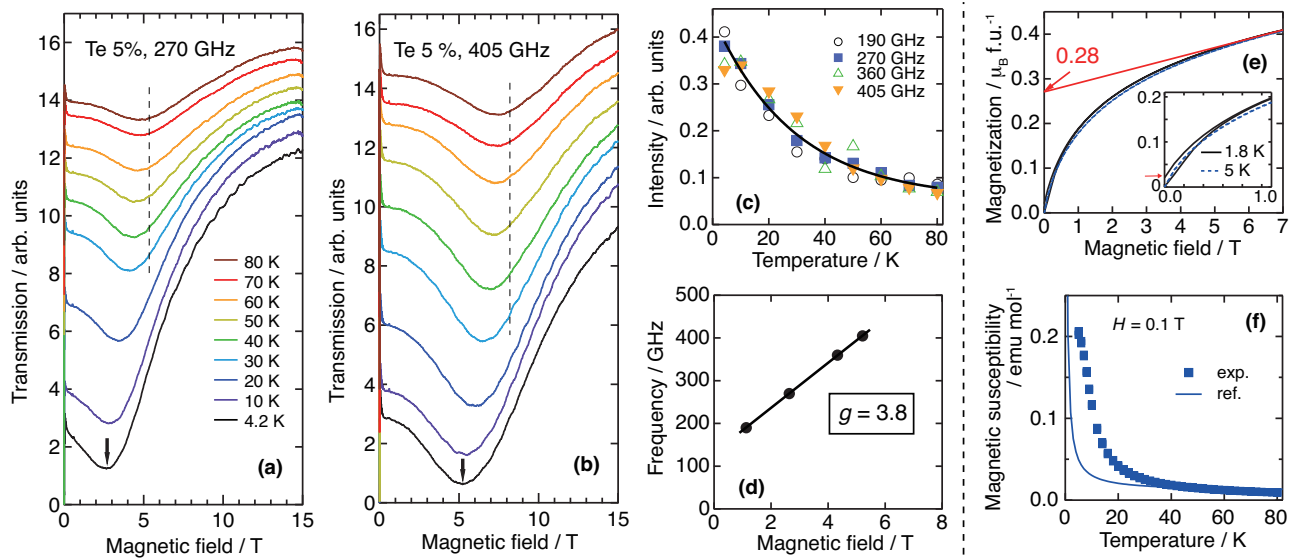}
\end{center}
\caption{\label{fig:ESR_mag} (Color online) (a)(b) ESR absorption spectra measured for LaCo$_{0.95}$Te$_{0.05}$O$_3$ between 4.2 K and 80 K. The arrows indicate the absorption bottoms at 4.2 K. The dotted vertical lines indicate the absorption bottom positions arising from thermally activated Co$^{3+{\rm (HS)}}$.~\cite{2002_Noguchi} 
(c) Temperature dependence of the integrated absorption intensity. The spectra were integrated between the absorption bottom and 15 T. The curve is a guide for the eye.
(d) Frequency-field diagram at 4.2 K. The line is a guide for the eye.
(e) Field dependence of the magnetization. The inset magnifies the low-field range.
(f) Temperature dependence of the magnetic susceptibility. The curve shows the calculated reference line, consisting of the Curie $1/T$ term obtained from $g=3.8$ and $j_{\rm eff}=1/2$ and the van Vleck constant term estimated from the range between 6--7 T in (e). }
\end{figure*}
%

Figures~\ref{fig:ESR_mag}(a) and \ref{fig:ESR_mag}(b) show the ESR absorption spectra measured for LaCo$_{0.95}$Te$_{0.05}$O$_3$ at typical frequencies. A concave absorption structure is observed in all the data. The absorption intensity is the strongest at the measured lowest temperature (4.2 K) and decreases with increasing temperature up to 80 K, as shown in Fig.~\ref{fig:ESR_mag}(c), indicating that the 4.2-K absorption corresponds to Co$^{2+}$. 
On the other hand, with increasing the temperature, the absorption bottom position approaches to the dotted vertical lines, corresponding to the bottom positions for the thermally activated Co$^{3+{\rm (HS)}}$.~\cite{2002_Noguchi} Hence, the high-temperature spectra will be interpreted as the superposition of the Co$^{2+}$ component and the thermally activated Co$^{3+{\rm (HS)}}$ component. 

Figure~\ref{fig:ESR_mag}(d) shows the magnetic-field dependence of absorption frequencies at 4.2 K. The slope gives the $g$ value of 3.8, confirming that the Co$^{2+}$ forms the HS state, rather than the LS state. 
We also remark that this Co$^{2+{\rm (HS)}}$ state differs from the often assumed spin-only model ($g=2$ and $S=3/2$). 

The XAS and ESR combined result is summarized as LaCo$^{3+{\rm(LS)}}_{0.80}$Co$^{2+{\rm(HS)}}_{0.15}$Te$^{6+}_{0.05}$O$_3$ in the lowest temperature range for $y=0.05$, in which the Co$^{2+{\rm (HS)}}$ state is described by $g=3.8$ and $j_{\rm eff}=1/2$. 
We compare this result to the magnetization data. 
(1) Figure~\ref{fig:ESR_mag}(e) shows the field dependence of magnetization measured at 1.8 K and 5 K. The paramagnetic Brillouin-like saturating behavior is observed with the maximum value 0.28$\mu_{\rm B}$ per formula unit (f.u.). This value corresponds to the 15{\%} concentration of Co$^{2+{\rm (HS)}}$, which coincides with the XAS and ESR result. 
(2) As shown in the inset in Fig.~\ref{fig:ESR_mag}(e), tiny finite remanent magnetization is observed at 1.8 K, meaning that a weak ferromagnetic correlation grows in the mainly paramagnetic state below 5 K. In fact, in the temperature dependence of magnetic susceptibility [Fig.~\ref{fig:ESR_mag}(f)], as the temperature decreases, the experimental values increase in advance of the calculated reference line obtained for the 15{\%} concentration of Co$^{2+{\rm (HS)}}$ in the ideally paramagnetic Curie law. 
Thus, the XAS, ESR, and magnetization results are consistently understood with each other.


\begin{table}[htbp]
\caption{\label{tab:kappa} Covalent-bonding parameter $\kappa$ for some systems.~\cite{1960_Low, 2002_Noguchi} The ratio of ionic radii ($r$) is denoted by $r_{{\rm doped-ion}}/r_{{\rm matrix-ion}}$; for example, $r_{\rm Co^{2+{\rm (HS)}}}/r_{\rm Mg^{2+}} = 1.0$, based on Shannon's radii.~\cite{1976_Shannon} }
\begin{ruledtabular}
\begin{tabular}{cccc}
Matrix & Doped ion & $\kappa$ & Ratio of ionic radius  \\
\hline
ZnF$_2$ & Co$^{2+{\rm (HS)}}$ & 0.9~\cite{1960_Low} & 1.0\\
MgO & Co$^{2+{\rm (HS)}}$ & 0.9~\cite{1960_Low} & 1.0\\
MgO & Fe$^{2+{\rm (HS)}}$ & 0.8~\cite{1960_Low} & 1.1\\
LaCoO$_3$ (LT) & Co$^{3+{\rm (HS)}}$ & 0.8~\cite{2002_Noguchi} & 1.1\\
LaCoO$_3$ (LT) & Co$^{2+{\rm (HS)}}$ & 0.5$^{\rm a}$ & 1.4\\
\end{tabular}
\end{ruledtabular}
\begin{flushleft}
$^{\rm a}$ This work.
\end{flushleft}
\end{table}

We further discuss $g=3.8$. This value is smaller than the typical value of 4.3, which is, for example, observed for Co$^{2+{\rm (HS)}}$ in MgO.~\cite{1958_Low} Low {\it et al.} correlated $g$ values to the metal-ligand bonding nature: $g = 10/3 + \kappa$ for the $d^7$ state [Co$^{2+{\rm (HS)}}$] and $g = 3 + (1/2)\kappa$ for the $d^6$ state [Fe$^{2+ {\rm (HS))}}$ and Co$^{3+ {\rm (HS))}}$], where $\kappa$ takes 1 in the ideal ionic-bonding case and decreases from 1 in accordance with the degree of the covalent-bonding nature.~\cite{1958_Low, 1960_Low} Table~\ref{tab:kappa} gives the $\kappa$ values estimated using these relations for some systems.~\cite{1960_Low, 2002_Noguchi} Among them, interestingly, $g=3.8$ gives a significantly low $\kappa=0.5$, indicating the highly covalent nature of Co$^{2+{\rm (HS)}}$--O$^{2-}$ bonding in the electron-doped LaCoO$_3$ system. Indeed, on increasing the ionic radius ratio of doped ions to matrix ions, which is also regarded as an index of chemical pressure and therefore covalency, the $\kappa$ value decreases. Thus, the Co$^{2+}$ ion, squeezed in the Co$^{3+{\rm(LS)}}$ matrix, forms the HS state but is probably accompanied with a strong covalent-boding nature. 

\subsection{Discussion of origins of spin-state blockade}
Thus far, we showed that the LT insulating range includes Co$^{3+{\rm (LS)}}$ and Co$^{2+{\rm (HS)}}$ and exhibits the predominantly high electrical resistivity compared to hole-doped systems. These results demonstrate the emergence of spin-state blockade phenomena. As its origin, first, the rule prohibiting Co$^{2+ {\rm (LS)}}$ is suggested, as mentioned in Introduction.~\cite{2004_Maignan, 2005_Taskin, 2009_Chang} However, in this subsection, we would like to point out that this is not the only origin or interpretation. The discussion is generally applicable to not only this-time LaCo$_{1-y}$Te$_{y}$O$_3$ but also $R$BaCo$_2$O$_{5.5+\delta}$ ($R$: rare earth), La$_{1.5}$Sr$_{0.5}$CoO$_4$, and other LaCoO$_3$-based electron-doped systems. 

The second probable origin is the difference of covalent-bonding strength. Goodenough presented the theory, in which the covalent-bonding strength provides a criterion to determine whether the outer electrons are best described by a localized-electron or a band/molecular-electron model.~\cite{1966_Goodenough} In his theory, LS states generate the extremely covalent $\sigma$*-orbital band connecting interatomic Co 3$d$-$e_g$ and Co 3$d$-$e_g$ through O 2$p$, whereas HS sites are localized. This suggests that the electron transfer from Co$^{2+{\rm (HS)}}$ to Co$^{3+{\rm (LS)}}$ is suppressed. Indeed, in turn, Co$^{4+}$ forms the LS state, or possibly the IS state like in SrCoO$_3$, in hole-doped systems that are much more conductive.~\cite{2008_Podlesnyak, 2011_Podlesnyak, 2011_Long} 

Our ESR study revealed the highly covalent Co$^{2+{\rm (HS)}}$. Combining with the aforementioned theory, this Co$^{2+{\rm (HS)}}$ is considered to be very close to the extremely covalent Co$^{2+{\rm (LS)}}$. Therefore, Co$^{2+{\rm (LS)}}$ that is very rare in oxides could be realized by a stimulus like pressure in future. 

The third possible origin is the large difference of ionic radii. The Co$^{2+{\rm (HS)}}$ is particularly large compared to the matrix Co$^{3+{\rm (LS)}}$ and others [$r_{{\rm Co^{2+{\rm (HS)}}}}=0.745$ {\AA}, $r_{{\rm Co}^{2+{\rm (LS)}}}=0.65$ {\AA}, $r_{{\rm Co}^{3+{\rm (HS)}}}=0.61$ {\AA}, and $r_{{\rm Co}^{3+{\rm (LS)}}}=0.545$ {\AA}].~\cite{1976_Shannon} Hence, the movement of Co$^{2+{\rm (HS)}}$ in the Co$^{3+{\rm (LS)}}$ matrix is accompanied by large local lattice deformation, which will costs the high activation energy based on elastic energy. This spin-state-lattice coupling is also expected to suppress the Co$^{2+{\rm (HS)}}$--Co$^{3+{\rm (LS)}}$ charge transfer. 

The other related factor may be the local structure around dopants, oxygen defects, and doped electrons. Thus, the spin-state blockade phenomena should be regarded as the total effect of these origins. 


%
\section{Conclusions}
%
Electron-doped LaCo$_{1-y}$Te$_{y}$O$_3$ was microscopically studied by measuring hard and soft x-ray absorption spectra and electron spin resonance. The LT insulating range includes Co$^{3+{\rm (LS)}}$ and Co$^{2+{\rm (HS)}}$, which coincides with the spin-state blockade in combination with the electron--hole asymmetry in electrical resistivity. 
Further, this Co$^{2+{\rm (HS)}}$ is considered to be accompanied with a significant covalent-bonding nature. 
We also discussed several origins of the spin-state blockade. 
Opportunities could be created for the further studies of electron-doped Co-spin-state physics, the realization of rare Co$^{2+{\rm (LS)}}$, and the development of materials and devices utilizing the spin-state blockade, such as sensors and gate switches with Co$^{3+{\rm (LS)}}$ creation/annihilation. 

\acknowledgments
We thank Mr. S. Kayamori for assisting with the sample characterization, Mr. S. Nara and Ms. E. P. Sinaga for assisting with the preliminary sample evaluation, Mr. K. Hashimoto for assisting with the preliminary ESR data analysis, Dr. N. Kimura for assisting with the SQUID measurements, and the NSRRC staff and students for their hospitality and assistance with the XAS measurements. The PF experiments have been performed under the approval of the Photon Factory Program Advisory Committee (Proposal No. 2009S2-008, 2012S2-005, and 2015PF-BL-19B). This study was financially supported by Grants-in-Aid for Young Scientists (B) (22740209 and 26800174) and Scientific Researches (S) (21224008) from MEXT, Japan. H. N. and Y. M. were financially supported by the Funding Program for World-leading Innovative R{\&}D in Science and Technology (FIRST).
\bibliography{LaCoTeO3_9c_arXiv}

\begin{thebibliography}{47}
\expandafter\ifx\csname natexlab\endcsname\relax\def\natexlab#1{#1}\fi
\expandafter\ifx\csname bibnamefont\endcsname\relax
  \def\bibnamefont#1{#1}\fi
\expandafter\ifx\csname bibfnamefont\endcsname\relax
  \def\bibfnamefont#1{#1}\fi
\expandafter\ifx\csname citenamefont\endcsname\relax
  \def\citenamefont#1{#1}\fi
\expandafter\ifx\csname url\endcsname\relax
  \def\url#1{\texttt{#1}}\fi
\expandafter\ifx\csname urlprefix\endcsname\relax\def\urlprefix{URL }\fi
\providecommand{\bibinfo}[2]{#2}
\providecommand{\eprint}[2][]{\url{#2}}

\bibitem[{\citenamefont{Anderson}(1972)}]{1972_Anderson}
\bibinfo{author}{\bibfnamefont{P.~W.} \bibnamefont{Anderson}},
  \bibinfo{journal}{Science} \textbf{\bibinfo{volume}{177}},
  \bibinfo{pages}{4047} (\bibinfo{year}{1972}).

\bibitem[{\citenamefont{\text{For example, } K.~Ono
  et~al.}(2002)\citenamefont{\text{For example, } K.~Ono, Austing, Tokura, and
  Tarucha}}]{2002_Ono}
\bibinfo{author}{\bibnamefont{\text{For example, } K.~Ono}},
  \bibinfo{author}{\bibfnamefont{D.~G.} \bibnamefont{Austing}},
  \bibinfo{author}{\bibfnamefont{Y.}~\bibnamefont{Tokura}}, \bibnamefont{and}
  \bibinfo{author}{\bibfnamefont{S.}~\bibnamefont{Tarucha}},
  \bibinfo{journal}{Science} \textbf{\bibinfo{volume}{297}},
  \bibinfo{pages}{1313} (\bibinfo{year}{2002}).

\bibitem[{\citenamefont{Maignan et~al.}(2004)\citenamefont{Maignan, Caignaert,
  Raveau, Khomskii, and Sawatzky}}]{2004_Maignan}
\bibinfo{author}{\bibfnamefont{A.}~\bibnamefont{Maignan}},
  \bibinfo{author}{\bibfnamefont{V.}~\bibnamefont{Caignaert}},
  \bibinfo{author}{\bibfnamefont{B.}~\bibnamefont{Raveau}},
  \bibinfo{author}{\bibfnamefont{D.}~\bibnamefont{Khomskii}}, \bibnamefont{and}
  \bibinfo{author}{\bibfnamefont{G.}~\bibnamefont{Sawatzky}},
  \bibinfo{journal}{Phys. Rev. Lett.} \textbf{\bibinfo{volume}{93}},
  \bibinfo{pages}{026401} (\bibinfo{year}{2004}).

\bibitem[{\citenamefont{Taskin and Ando}(2005)}]{2005_Taskin}
\bibinfo{author}{\bibfnamefont{A.~A.} \bibnamefont{Taskin}} \bibnamefont{and}
  \bibinfo{author}{\bibfnamefont{Y.}~\bibnamefont{Ando}},
  \bibinfo{journal}{Phys. Rev. Lett.} \textbf{\bibinfo{volume}{95}},
  \bibinfo{pages}{176603} (\bibinfo{year}{2005}).

\bibitem[{\citenamefont{{C. F. Chang, Z. Hu, H. Wu, T. Burnus, N. Hollmann, M.
  Benomar, T. Lorenz, A. Tanaka, H.-J. Lin, H. H. Hsieh, C. T. Chen, and L. H.
  Tjeng}}(2009)}]{2009_Chang}
\bibinfo{author}{\bibnamefont{{C. F. Chang, Z. Hu, H. Wu, T. Burnus, N.
  Hollmann, M. Benomar, T. Lorenz, A. Tanaka, H.-J. Lin, H. H. Hsieh, C. T.
  Chen, and L. H. Tjeng}}}, \bibinfo{journal}{Phys. Rev. Lett.}
  \textbf{\bibinfo{volume}{102}}, \bibinfo{pages}{116401}
  (\bibinfo{year}{2009}).

\bibitem[{\citenamefont{Korotin et~al.}(1996)\citenamefont{Korotin, Ezhov,
  Solovyev, Anisimov, Khomskii, and Sawatzky}}]{1996_Korotin}
\bibinfo{author}{\bibfnamefont{M.~A.} \bibnamefont{Korotin}},
  \bibinfo{author}{\bibfnamefont{S.~Y.} \bibnamefont{Ezhov}},
  \bibinfo{author}{\bibfnamefont{I.~V.} \bibnamefont{Solovyev}},
  \bibinfo{author}{\bibfnamefont{V.~I.} \bibnamefont{Anisimov}},
  \bibinfo{author}{\bibfnamefont{D.~I.} \bibnamefont{Khomskii}},
  \bibnamefont{and} \bibinfo{author}{\bibfnamefont{G.~A.}
  \bibnamefont{Sawatzky}}, \bibinfo{journal}{Phys. Rev. B}
  \textbf{\bibinfo{volume}{54}}, \bibinfo{pages}{5309} (\bibinfo{year}{1996}).

\bibitem[{\citenamefont{Sugano et~al.}(Academic, New York,
  1970)\citenamefont{Sugano, Tanabe, and Kamimura}}]{1970_Sugano}
\bibinfo{author}{\bibfnamefont{S.}~\bibnamefont{Sugano}},
  \bibinfo{author}{\bibfnamefont{Y.}~\bibnamefont{Tanabe}}, \bibnamefont{and}
  \bibinfo{author}{\bibfnamefont{H.}~\bibnamefont{Kamimura}},
  \bibinfo{journal}{{\it Multiplets of Transition-Metal Ions in Crystals}}
  (\bibinfo{year}{Academic, New York, 1970}).

\bibitem[{\citenamefont{Heikes et~al.}(1964)\citenamefont{Heikes, Miller, and
  Mazelsky}}]{1964_Heikes}
\bibinfo{author}{\bibfnamefont{R.~R.} \bibnamefont{Heikes}},
  \bibinfo{author}{\bibfnamefont{R.~C.} \bibnamefont{Miller}},
  \bibnamefont{and} \bibinfo{author}{\bibfnamefont{R.}~\bibnamefont{Mazelsky}},
  \bibinfo{journal}{Physica} \textbf{\bibinfo{volume}{30}},
  \bibinfo{pages}{1600} (\bibinfo{year}{1964}).

\bibitem[{\citenamefont{Bhide et~al.}(1972)\citenamefont{Bhide, Rajoria, Rao,
  and Rao}}]{1972_Bhide}
\bibinfo{author}{\bibfnamefont{V.~G.} \bibnamefont{Bhide}},
  \bibinfo{author}{\bibfnamefont{D.~S.} \bibnamefont{Rajoria}},
  \bibinfo{author}{\bibfnamefont{G.~R.} \bibnamefont{Rao}}, \bibnamefont{and}
  \bibinfo{author}{\bibfnamefont{C.~N.~R.} \bibnamefont{Rao}},
  \bibinfo{journal}{Phys. Rev. B} \textbf{\bibinfo{volume}{6}},
  \bibinfo{pages}{1021} (\bibinfo{year}{1972}).

\bibitem[{\citenamefont{{M. W. Haverkort, Z. Hu, J. C. Cezar, T. Burnus, H.
  Hartmann, M. Reuther, C. Zobel, T. Lorenz, A. Tanaka, N. B. Brookes, H. H.
  Hsieh, H.-J. Lin, C. T. Chen, and L. H. Tjeng}}(2006)}]{2006_Haverkort}
\bibinfo{author}{\bibnamefont{{M. W. Haverkort, Z. Hu, J. C. Cezar, T. Burnus,
  H. Hartmann, M. Reuther, C. Zobel, T. Lorenz, A. Tanaka, N. B. Brookes, H. H.
  Hsieh, H.-J. Lin, C. T. Chen, and L. H. Tjeng}}}, \bibinfo{journal}{Phys.
  Rev. Lett.} \textbf{\bibinfo{volume}{97}}, \bibinfo{pages}{176405}
  (\bibinfo{year}{2006}).

\bibitem[{\citenamefont{Sato et~al.}(2009)\citenamefont{Sato, Matsuo, Kindo,
  Kobayashi, and Asai}}]{2009_Sato}
\bibinfo{author}{\bibfnamefont{K.}~\bibnamefont{Sato}},
  \bibinfo{author}{\bibfnamefont{A.}~\bibnamefont{Matsuo}},
  \bibinfo{author}{\bibfnamefont{K.}~\bibnamefont{Kindo}},
  \bibinfo{author}{\bibfnamefont{Y.}~\bibnamefont{Kobayashi}},
  \bibnamefont{and} \bibinfo{author}{\bibfnamefont{K.}~\bibnamefont{Asai}},
  \bibinfo{journal}{J. Phys. Soc. Jpn} \textbf{\bibinfo{volume}{78}},
  \bibinfo{pages}{093702} (\bibinfo{year}{2009}).

\bibitem[{\citenamefont{Asai et~al.}(1997)\citenamefont{Asai, Yokokura, Suzuki,
  Naka, Matsumoto, Takahashi, M{\^o}ri, and Kohn}}]{1997_Asai}
\bibinfo{author}{\bibfnamefont{K.}~\bibnamefont{Asai}},
  \bibinfo{author}{\bibfnamefont{O.}~\bibnamefont{Yokokura}},
  \bibinfo{author}{\bibfnamefont{M.}~\bibnamefont{Suzuki}},
  \bibinfo{author}{\bibfnamefont{T.}~\bibnamefont{Naka}},
  \bibinfo{author}{\bibfnamefont{T.}~\bibnamefont{Matsumoto}},
  \bibinfo{author}{\bibfnamefont{H.}~\bibnamefont{Takahashi}},
  \bibinfo{author}{\bibfnamefont{N.}~\bibnamefont{M{\^o}ri}}, \bibnamefont{and}
  \bibinfo{author}{\bibfnamefont{K.}~\bibnamefont{Kohn}}, \bibinfo{journal}{J.
  Phys. Soc. Jpn} \textbf{\bibinfo{volume}{66}}, \bibinfo{pages}{967}
  (\bibinfo{year}{1997}).

\bibitem[{\citenamefont{Yamaguchi et~al.}(1996)\citenamefont{Yamaguchi,
  Okimoto, Taniguchi, and Tokura}}]{1996_Yamaguchi}
\bibinfo{author}{\bibfnamefont{S.}~\bibnamefont{Yamaguchi}},
  \bibinfo{author}{\bibfnamefont{Y.}~\bibnamefont{Okimoto}},
  \bibinfo{author}{\bibfnamefont{H.}~\bibnamefont{Taniguchi}},
  \bibnamefont{and} \bibinfo{author}{\bibfnamefont{Y.}~\bibnamefont{Tokura}},
  \bibinfo{journal}{Phys. Rev. B} \textbf{\bibinfo{volume}{53}},
  \bibinfo{pages}{R2926} (\bibinfo{year}{1996}).

\bibitem[{\citenamefont{Gerthsen and H$\ddot{\rm
  a}$rdtl}(1962)}]{1962_Gerthsen}
\bibinfo{author}{\bibfnamefont{V.~P.} \bibnamefont{Gerthsen}} \bibnamefont{and}
  \bibinfo{author}{\bibfnamefont{K.~U.} \bibnamefont{H$\ddot{\rm a}$rdtl}},
  \bibinfo{journal}{Z. Naturforschg. (German)} \textbf{\bibinfo{volume}{17 a}},
  \bibinfo{pages}{514} (\bibinfo{year}{1962}).

\bibitem[{\citenamefont{Zheng et~al.}(2008)\citenamefont{Zheng, Zhu, Song, and
  Sun}}]{2008_Zheng}
\bibinfo{author}{\bibfnamefont{G.~H.} \bibnamefont{Zheng}},
  \bibinfo{author}{\bibfnamefont{X.~B.} \bibnamefont{Zhu}},
  \bibinfo{author}{\bibfnamefont{W.~H.} \bibnamefont{Song}}, \bibnamefont{and}
  \bibinfo{author}{\bibfnamefont{Y.~P.} \bibnamefont{Sun}},
  \bibinfo{journal}{J. Appl. Phys.} \textbf{\bibinfo{volume}{103}},
  \bibinfo{pages}{013906} (\bibinfo{year}{2008}).

\bibitem[{\citenamefont{Jir$\acute{\rm a}$k
  et~al.}(2008)\citenamefont{Jir$\acute{\rm a}$k, Hejtm$\acute{\rm a}$nek,
  Kn$\acute{\rm \i}$$\check{z}$ek, and Veverka}}]{2008_Jirak}
\bibinfo{author}{\bibfnamefont{Z.}~\bibnamefont{Jir$\acute{\rm a}$k}},
  \bibinfo{author}{\bibfnamefont{J.}~\bibnamefont{Hejtm$\acute{\rm a}$nek}},
  \bibinfo{author}{\bibfnamefont{K.}~\bibnamefont{Kn$\acute{\rm
  \i}$$\check{z}$ek}}, \bibnamefont{and}
  \bibinfo{author}{\bibfnamefont{M.}~\bibnamefont{Veverka}},
  \bibinfo{journal}{Phys. Rev. B} \textbf{\bibinfo{volume}{78}},
  \bibinfo{pages}{014432} (\bibinfo{year}{2008}).

\bibitem[{\citenamefont{$\acute{\rm A}$lvarez Serrano
  et~al.}(2008)\citenamefont{$\acute{\rm A}$lvarez Serrano, Cuello,
  L$\acute{\rm o}$pez, Jim$\acute{\rm e}$nez-L$\acute{\rm o}$pez, Pico,
  Rodr$\acute{\rm \i}$guez-Castell$\acute{\rm o}$n, Rodr$\acute{\rm \i}$guez,
  and Veiga}}]{2008_Alvarez-Serrano}
\bibinfo{author}{\bibfnamefont{I.}~\bibnamefont{$\acute{\rm A}$lvarez
  Serrano}}, \bibinfo{author}{\bibfnamefont{G.~J.} \bibnamefont{Cuello}},
  \bibinfo{author}{\bibfnamefont{M.~L.} \bibnamefont{L$\acute{\rm o}$pez}},
  \bibinfo{author}{\bibfnamefont{A.}~\bibnamefont{Jim$\acute{\rm
  e}$nez-L$\acute{\rm o}$pez}},
  \bibinfo{author}{\bibfnamefont{C.}~\bibnamefont{Pico}},
  \bibinfo{author}{\bibfnamefont{E.}~\bibnamefont{Rodr$\acute{\rm
  \i}$guez-Castell$\acute{\rm o}$n}},
  \bibinfo{author}{\bibfnamefont{E.}~\bibnamefont{Rodr$\acute{\rm \i}$guez}},
  \bibnamefont{and} \bibinfo{author}{\bibfnamefont{M.~L.} \bibnamefont{Veiga}},
  \bibinfo{journal}{J. Phys. D: Appl. Phys.} \textbf{\bibinfo{volume}{41}},
  \bibinfo{pages}{195001} (\bibinfo{year}{2008}).

\bibitem[{\citenamefont{Hejtm$\acute{\rm a}$neka
  et~al.}(2008)\citenamefont{Hejtm$\acute{\rm a}$neka, Jir$\acute{\rm a}$k,
  Kn$\acute{\rm \i}$$\check{\rm z}$ek, Mary$\check{\rm s}$ko, Veverka, and
  Autret}}]{2008_Hejtmanek}
\bibinfo{author}{\bibfnamefont{J.}~\bibnamefont{Hejtm$\acute{\rm a}$neka}},
  \bibinfo{author}{\bibfnamefont{Z.}~\bibnamefont{Jir$\acute{\rm a}$k}},
  \bibinfo{author}{\bibfnamefont{K.}~\bibnamefont{Kn$\acute{\rm \i}$$\check{\rm
  z}$ek}}, \bibinfo{author}{\bibfnamefont{M.}~\bibnamefont{Mary$\check{\rm
  s}$ko}}, \bibinfo{author}{\bibfnamefont{M.}~\bibnamefont{Veverka}},
  \bibnamefont{and} \bibinfo{author}{\bibfnamefont{C.}~\bibnamefont{Autret}},
  \bibinfo{journal}{J. Magn. Magn. Mater.} \textbf{\bibinfo{volume}{78}},
  \bibinfo{pages}{014432} (\bibinfo{year}{2008}).

\bibitem[{\citenamefont{Nakatsugawa and Iguchi}(1999)}]{1999_Nakatsugawa}
\bibinfo{author}{\bibfnamefont{H.}~\bibnamefont{Nakatsugawa}} \bibnamefont{and}
  \bibinfo{author}{\bibfnamefont{E.}~\bibnamefont{Iguchi}},
  \bibinfo{journal}{J. Phys.:Condens. Matter} \textbf{\bibinfo{volume}{11}},
  \bibinfo{pages}{1711} (\bibinfo{year}{1999}).

\bibitem[{\citenamefont{Augustinsk$\acute{{\rm y}}$
  et~al.}(2013)\citenamefont{Augustinsk$\acute{{\rm y}}$, K$\check{{\rm
  r}}$$\acute{{\rm a}}$pek, and Kune$\check{{\rm s}}$}}]{2013_Augustinsky}
\bibinfo{author}{\bibfnamefont{P.}~\bibnamefont{Augustinsk$\acute{{\rm y}}$}},
  \bibinfo{author}{\bibfnamefont{V.}~\bibnamefont{K$\check{{\rm
  r}}$$\acute{{\rm a}}$pek}}, \bibnamefont{and}
  \bibinfo{author}{\bibfnamefont{J.}~\bibnamefont{Kune$\check{{\rm s}}$}},
  \bibinfo{journal}{Phys. Rev. Lett.} \textbf{\bibinfo{volume}{110}},
  \bibinfo{pages}{267204} (\bibinfo{year}{2013}).

\bibitem[{\citenamefont{Kanamori}(1957)}]{1957_Kanamori_a}
\bibinfo{author}{\bibfnamefont{J.}~\bibnamefont{Kanamori}},
  \bibinfo{journal}{Prog. Theor. Phys.} \textbf{\bibinfo{volume}{17}},
  \bibinfo{pages}{177} (\bibinfo{year}{1957}).

\bibitem[{\citenamefont{Noguchi et~al.}(2002)\citenamefont{Noguchi, Kawamata,
  Okuda, Nojiri, and Motokawa}}]{2002_Noguchi}
\bibinfo{author}{\bibfnamefont{S.}~\bibnamefont{Noguchi}},
  \bibinfo{author}{\bibfnamefont{S.}~\bibnamefont{Kawamata}},
  \bibinfo{author}{\bibfnamefont{K.}~\bibnamefont{Okuda}},
  \bibinfo{author}{\bibfnamefont{H.}~\bibnamefont{Nojiri}}, \bibnamefont{and}
  \bibinfo{author}{\bibfnamefont{M.}~\bibnamefont{Motokawa}},
  \bibinfo{journal}{Phys. Rev. B} \textbf{\bibinfo{volume}{66}},
  \bibinfo{pages}{094404} (\bibinfo{year}{2002}).

\bibitem[{\citenamefont{Podlesnyak et~al.}(2006)\citenamefont{Podlesnyak,
  Streule, Mesot, Medarde, Pomjakushina, Conder, Tanaka, Haverkort, and
  Khomskii}}]{2006_Podlesnyak}
\bibinfo{author}{\bibfnamefont{A.}~\bibnamefont{Podlesnyak}},
  \bibinfo{author}{\bibfnamefont{S.}~\bibnamefont{Streule}},
  \bibinfo{author}{\bibfnamefont{J.}~\bibnamefont{Mesot}},
  \bibinfo{author}{\bibfnamefont{M.}~\bibnamefont{Medarde}},
  \bibinfo{author}{\bibfnamefont{E.}~\bibnamefont{Pomjakushina}},
  \bibinfo{author}{\bibfnamefont{K.}~\bibnamefont{Conder}},
  \bibinfo{author}{\bibfnamefont{A.}~\bibnamefont{Tanaka}},
  \bibinfo{author}{\bibfnamefont{M.~W.} \bibnamefont{Haverkort}},
  \bibnamefont{and} \bibinfo{author}{\bibfnamefont{D.~I.}
  \bibnamefont{Khomskii}}, \bibinfo{journal}{Phys. Rev. Lett.}
  \textbf{\bibinfo{volume}{97}}, \bibinfo{pages}{247208}
  (\bibinfo{year}{2006}).

\bibitem[{\citenamefont{Kriener et~al.}(2004)\citenamefont{Kriener, Zobel,
  Reichl, Baier, Cwik, Berggold, Kierspel, Zabara, Freimuth, and
  Lorenz}}]{2004_Kriener}
\bibinfo{author}{\bibfnamefont{M.}~\bibnamefont{Kriener}},
  \bibinfo{author}{\bibfnamefont{C.}~\bibnamefont{Zobel}},
  \bibinfo{author}{\bibfnamefont{A.}~\bibnamefont{Reichl}},
  \bibinfo{author}{\bibfnamefont{J.}~\bibnamefont{Baier}},
  \bibinfo{author}{\bibfnamefont{M.}~\bibnamefont{Cwik}},
  \bibinfo{author}{\bibfnamefont{K.}~\bibnamefont{Berggold}},
  \bibinfo{author}{\bibfnamefont{H.}~\bibnamefont{Kierspel}},
  \bibinfo{author}{\bibfnamefont{O.}~\bibnamefont{Zabara}},
  \bibinfo{author}{\bibfnamefont{A.}~\bibnamefont{Freimuth}}, \bibnamefont{and}
  \bibinfo{author}{\bibfnamefont{T.}~\bibnamefont{Lorenz}},
  \bibinfo{journal}{Phys. Rev. B} \textbf{\bibinfo{volume}{69}},
  \bibinfo{pages}{094417} (\bibinfo{year}{2004}).

\bibitem[{\citenamefont{Wu and Leighton}(2003)}]{2003_Wu}
\bibinfo{author}{\bibfnamefont{J.}~\bibnamefont{Wu}} \bibnamefont{and}
  \bibinfo{author}{\bibfnamefont{C.}~\bibnamefont{Leighton}},
  \bibinfo{journal}{Phys. Rev. B} \textbf{\bibinfo{volume}{67}},
  \bibinfo{pages}{174408} (\bibinfo{year}{2003}).

\bibitem[{\citenamefont{Long et~al.}(2011)\citenamefont{Long, Kaneko, Ishiwata,
  Taguchi, and Tokura}}]{2011_Long}
\bibinfo{author}{\bibfnamefont{Y.}~\bibnamefont{Long}},
  \bibinfo{author}{\bibfnamefont{Y.}~\bibnamefont{Kaneko}},
  \bibinfo{author}{\bibfnamefont{S.}~\bibnamefont{Ishiwata}},
  \bibinfo{author}{\bibfnamefont{Y.}~\bibnamefont{Taguchi}}, \bibnamefont{and}
  \bibinfo{author}{\bibfnamefont{Y.}~\bibnamefont{Tokura}},
  \bibinfo{journal}{J. Phys.: Condens. Matter} \textbf{\bibinfo{volume}{23}},
  \bibinfo{pages}{245601} (\bibinfo{year}{2011}).

\bibitem[{\citenamefont{Tabata and Kido}(1989)}]{1989_Tabata}
\bibinfo{author}{\bibfnamefont{K.}~\bibnamefont{Tabata}} \bibnamefont{and}
  \bibinfo{author}{\bibfnamefont{H.}~\bibnamefont{Kido}},
  \bibinfo{journal}{Phys. Stat. Sol. (a)} \textbf{\bibinfo{volume}{111}},
  \bibinfo{pages}{K105} (\bibinfo{year}{1989}).

\bibitem[{\citenamefont{Fuchs et~al.}(2005)\citenamefont{Fuchs, Schweiss,
  Adelmann, Schwarz, and Schneider}}]{2005_Fuchs}
\bibinfo{author}{\bibfnamefont{D.}~\bibnamefont{Fuchs}},
  \bibinfo{author}{\bibfnamefont{P.}~\bibnamefont{Schweiss}},
  \bibinfo{author}{\bibfnamefont{P.}~\bibnamefont{Adelmann}},
  \bibinfo{author}{\bibfnamefont{T.}~\bibnamefont{Schwarz}}, \bibnamefont{and}
  \bibinfo{author}{\bibfnamefont{R.}~\bibnamefont{Schneider}},
  \bibinfo{journal}{Phys. Rev. B} \textbf{\bibinfo{volume}{72}},
  \bibinfo{pages}{014466} (\bibinfo{year}{2005}).

\bibitem[{\citenamefont{Tabata and Kohiki}(1987)}]{1987_Tabata}
\bibinfo{author}{\bibfnamefont{K.}~\bibnamefont{Tabata}} \bibnamefont{and}
  \bibinfo{author}{\bibfnamefont{S.}~\bibnamefont{Kohiki}},
  \bibinfo{journal}{J. Mat. Sci.} \textbf{\bibinfo{volume}{22}},
  \bibinfo{pages}{3781} (\bibinfo{year}{1987}).

\bibitem[{\citenamefont{Cairns et~al.}(2005)\citenamefont{Cairns, Reaney,
  Zheng, Iddles, and Price}}]{2005_Cairns}
\bibinfo{author}{\bibfnamefont{D.~L.} \bibnamefont{Cairns}},
  \bibinfo{author}{\bibfnamefont{I.~M.} \bibnamefont{Reaney}},
  \bibinfo{author}{\bibfnamefont{H.}~\bibnamefont{Zheng}},
  \bibinfo{author}{\bibfnamefont{D.}~\bibnamefont{Iddles}}, \bibnamefont{and}
  \bibinfo{author}{\bibfnamefont{T.}~\bibnamefont{Price}}, \bibinfo{journal}{J.
  Euro. Ceramic Soc.} \textbf{\bibinfo{volume}{25}}, \bibinfo{pages}{433}
  (\bibinfo{year}{2005}).

\bibitem[{\citenamefont{Robert et~al.}(2010)\citenamefont{Robert, Logvinovich,
  Aguirre, Ebbinghaus, Bocher, Tome$\check{\rm s}$, and
  Weidenkaff}}]{2010_Robert}
\bibinfo{author}{\bibfnamefont{R.}~\bibnamefont{Robert}},
  \bibinfo{author}{\bibfnamefont{D.}~\bibnamefont{Logvinovich}},
  \bibinfo{author}{\bibfnamefont{M.~H.} \bibnamefont{Aguirre}},
  \bibinfo{author}{\bibfnamefont{S.~G.} \bibnamefont{Ebbinghaus}},
  \bibinfo{author}{\bibfnamefont{L.}~\bibnamefont{Bocher}},
  \bibinfo{author}{\bibfnamefont{P.}~\bibnamefont{Tome$\check{\rm s}$}},
  \bibnamefont{and}
  \bibinfo{author}{\bibfnamefont{A.}~\bibnamefont{Weidenkaff}},
  \bibinfo{journal}{Acta Mater.} \textbf{\bibinfo{volume}{58}},
  \bibinfo{pages}{680} (\bibinfo{year}{2010}).

\bibitem[{\citenamefont{Shannon}(1976)}]{1976_Shannon}
\bibinfo{author}{\bibfnamefont{R.~D.} \bibnamefont{Shannon}},
  \bibinfo{journal}{Acta Cryst.} \textbf{\bibinfo{volume}{A32}},
  \bibinfo{pages}{751} (\bibinfo{year}{1976}).

\bibitem[{\citenamefont{Pinta et~al.}(2007)\citenamefont{Pinta, Fuchs,
  Pellegrin, Adelmann, Mangold, and Schuppler}}]{2007_Pinta}
\bibinfo{author}{\bibfnamefont{C.}~\bibnamefont{Pinta}},
  \bibinfo{author}{\bibfnamefont{D.}~\bibnamefont{Fuchs}},
  \bibinfo{author}{\bibfnamefont{E.}~\bibnamefont{Pellegrin}},
  \bibinfo{author}{\bibfnamefont{P.}~\bibnamefont{Adelmann}},
  \bibinfo{author}{\bibfnamefont{S.}~\bibnamefont{Mangold}}, \bibnamefont{and}
  \bibinfo{author}{\bibfnamefont{S.}~\bibnamefont{Schuppler}},
  \bibinfo{journal}{J. Low Temp. Phys.} \textbf{\bibinfo{volume}{147}},
  \bibinfo{pages}{314} (\bibinfo{year}{2007}).

\bibitem[{\citenamefont{Oishi et~al.}(2009)\citenamefont{Oishi, Yonemura,
  Nishimaki, Torii, Hoshikawa, Ishigaki, Morishima, Mori, and
  Kamiyama}}]{Z_2009}
\bibinfo{author}{\bibfnamefont{R.}~\bibnamefont{Oishi}},
  \bibinfo{author}{\bibfnamefont{M.}~\bibnamefont{Yonemura}},
  \bibinfo{author}{\bibfnamefont{Y.}~\bibnamefont{Nishimaki}},
  \bibinfo{author}{\bibfnamefont{S.}~\bibnamefont{Torii}},
  \bibinfo{author}{\bibfnamefont{A.}~\bibnamefont{Hoshikawa}},
  \bibinfo{author}{\bibfnamefont{T.}~\bibnamefont{Ishigaki}},
  \bibinfo{author}{\bibfnamefont{T.}~\bibnamefont{Morishima}},
  \bibinfo{author}{\bibfnamefont{K.}~\bibnamefont{Mori}}, \bibnamefont{and}
  \bibinfo{author}{\bibfnamefont{T.}~\bibnamefont{Kamiyama}},
  \bibinfo{journal}{Nucl. Instr. Meth. Phys. Res.}
  \textbf{\bibinfo{volume}{A600}}, \bibinfo{pages}{94} (\bibinfo{year}{2009}).

\bibitem[{\citenamefont{Oishi-Tomiyasu
  et~al.}(2012)\citenamefont{Oishi-Tomiyasu, Yonemura, Morishima, Hoshikawa,
  Torii, Ishigaki, and Kamiyama}}]{Z_2012}
\bibinfo{author}{\bibfnamefont{R.}~\bibnamefont{Oishi-Tomiyasu}},
  \bibinfo{author}{\bibfnamefont{M.}~\bibnamefont{Yonemura}},
  \bibinfo{author}{\bibfnamefont{T.}~\bibnamefont{Morishima}},
  \bibinfo{author}{\bibfnamefont{A.}~\bibnamefont{Hoshikawa}},
  \bibinfo{author}{\bibfnamefont{S.}~\bibnamefont{Torii}},
  \bibinfo{author}{\bibfnamefont{T.}~\bibnamefont{Ishigaki}}, \bibnamefont{and}
  \bibinfo{author}{\bibfnamefont{T.}~\bibnamefont{Kamiyama}},
  \bibinfo{journal}{J. Appl. Cryst.} \textbf{\bibinfo{volume}{45}},
  \bibinfo{pages}{299} (\bibinfo{year}{2012}).

\bibitem[{\citenamefont{Radaelli and Cheong}(2002)}]{2002_Radaelli}
\bibinfo{author}{\bibfnamefont{P.~G.} \bibnamefont{Radaelli}} \bibnamefont{and}
  \bibinfo{author}{\bibfnamefont{S.-W.} \bibnamefont{Cheong}},
  \bibinfo{journal}{Phys. Rev. B} \textbf{\bibinfo{volume}{66}},
  \bibinfo{pages}{094408} (\bibinfo{year}{2002}).

\bibitem[{\citenamefont{Politova and Venevtsev}(1975)}]{1975_Politova}
\bibinfo{author}{\bibfnamefont{E.~D.} \bibnamefont{Politova}} \bibnamefont{and}
  \bibinfo{author}{\bibfnamefont{Y.~N.} \bibnamefont{Venevtsev}},
  \bibinfo{journal}{Mat. Res. Bull.} \textbf{\bibinfo{volume}{10}},
  \bibinfo{pages}{319} (\bibinfo{year}{1975}).

\bibitem[{\citenamefont{Tan et~al.}(2003)\citenamefont{Tan, Dai, Duan, Zhou,
  Lu, and Chen}}]{2003_Tan}
\bibinfo{author}{\bibfnamefont{G.~T.} \bibnamefont{Tan}},
  \bibinfo{author}{\bibfnamefont{S.~Y.} \bibnamefont{Dai}},
  \bibinfo{author}{\bibfnamefont{P.}~\bibnamefont{Duan}},
  \bibinfo{author}{\bibfnamefont{Y.~L.} \bibnamefont{Zhou}},
  \bibinfo{author}{\bibfnamefont{H.~B.} \bibnamefont{Lu}}, \bibnamefont{and}
  \bibinfo{author}{\bibfnamefont{Z.~H.} \bibnamefont{Chen}},
  \bibinfo{journal}{J. Appl. Phys.} \textbf{\bibinfo{volume}{93}},
  \bibinfo{pages}{5480} (\bibinfo{year}{2003}).

\bibitem[{\citenamefont{Hu et~al.}(2012)\citenamefont{Hu, Wu, Koethe, Barilo,
  Shiryaev, Bychkov, Sch$\ddot{\rm u}$sler-Langeheine, Lorenz, Tanaka, Hsieh
  et~al.}}]{2012_Hu}
\bibinfo{author}{\bibfnamefont{Z.}~\bibnamefont{Hu}},
  \bibinfo{author}{\bibfnamefont{H.}~\bibnamefont{Wu}},
  \bibinfo{author}{\bibfnamefont{T.~C.} \bibnamefont{Koethe}},
  \bibinfo{author}{\bibfnamefont{S.~N.} \bibnamefont{Barilo}},
  \bibinfo{author}{\bibfnamefont{S.~V.} \bibnamefont{Shiryaev}},
  \bibinfo{author}{\bibfnamefont{G.~L.} \bibnamefont{Bychkov}},
  \bibinfo{author}{\bibfnamefont{C.}~\bibnamefont{Sch$\ddot{\rm
  u}$sler-Langeheine}},
  \bibinfo{author}{\bibfnamefont{T.}~\bibnamefont{Lorenz}},
  \bibinfo{author}{\bibfnamefont{A.}~\bibnamefont{Tanaka}},
  \bibinfo{author}{\bibfnamefont{H.~H.} \bibnamefont{Hsieh}},
  \bibnamefont{et~al.}, \bibinfo{journal}{New J. Phys.}
  \textbf{\bibinfo{volume}{14}}, \bibinfo{pages}{123025}
  (\bibinfo{year}{2012}).

\bibitem[{\citenamefont{Low}(1958)}]{1958_Low}
\bibinfo{author}{\bibfnamefont{W.}~\bibnamefont{Low}}, \bibinfo{journal}{Phys.
  Rev} \textbf{\bibinfo{volume}{109}}, \bibinfo{pages}{256}
  (\bibinfo{year}{1958}).

\bibitem[{\citenamefont{Lines}(1963)}]{1963_Lines}
\bibinfo{author}{\bibfnamefont{M.~E.} \bibnamefont{Lines}},
  \bibinfo{journal}{Phys. Rev} \textbf{\bibinfo{volume}{131}},
  \bibinfo{pages}{546} (\bibinfo{year}{1963}).

\bibitem[{\citenamefont{Dance et~al.}(1994)\citenamefont{Dance, Boireau,
  Lirzin, and Lestienne}}]{1994_Dance}
\bibinfo{author}{\bibfnamefont{J.~M.} \bibnamefont{Dance}},
  \bibinfo{author}{\bibfnamefont{A.}~\bibnamefont{Boireau}},
  \bibinfo{author}{\bibfnamefont{A.~L.} \bibnamefont{Lirzin}},
  \bibnamefont{and}
  \bibinfo{author}{\bibfnamefont{B.}~\bibnamefont{Lestienne}},
  \bibinfo{journal}{Solid State Commun.} \textbf{\bibinfo{volume}{91}},
  \bibinfo{pages}{475} (\bibinfo{year}{1994}).

\bibitem[{\citenamefont{Hoffmann et~al.}(2013)\citenamefont{Hoffmann, Goslar,
  and Lijewski}}]{2013_Hoffmann}
\bibinfo{author}{\bibfnamefont{S.~K.} \bibnamefont{Hoffmann}},
  \bibinfo{author}{\bibfnamefont{J.}~\bibnamefont{Goslar}}, \bibnamefont{and}
  \bibinfo{author}{\bibfnamefont{S.}~\bibnamefont{Lijewski}},
  \bibinfo{journal}{Appl. Mag. Res.} \textbf{\bibinfo{volume}{44}},
  \bibinfo{pages}{817} (\bibinfo{year}{2013}).

\bibitem[{\citenamefont{Low and Weger}(1960)}]{1960_Low}
\bibinfo{author}{\bibfnamefont{W.}~\bibnamefont{Low}} \bibnamefont{and}
  \bibinfo{author}{\bibfnamefont{M.}~\bibnamefont{Weger}},
  \bibinfo{journal}{Phys. Rev}  (\bibinfo{year}{1960}),
  \bibinfo{note}{\textbf{118}, 1119; \textbf{118}, 1130; \textbf{120}, 2277}.

\bibitem[{\citenamefont{Goodenough}(1966)}]{1966_Goodenough}
\bibinfo{author}{\bibfnamefont{J.~B.} \bibnamefont{Goodenough}},
  \bibinfo{journal}{J. Appl. Phys.} \textbf{\bibinfo{volume}{37}},
  \bibinfo{pages}{1415} (\bibinfo{year}{1966}).

\bibitem[{\citenamefont{Podlesnyak et~al.}(2008)\citenamefont{Podlesnyak,
  Russina, Furrer, Alfonsov, Vavilova, Kataev, B{\"u}chner, Str{\"a}ssle,
  Pomjakushina, Conder et~al.}}]{2008_Podlesnyak}
\bibinfo{author}{\bibfnamefont{A.}~\bibnamefont{Podlesnyak}},
  \bibinfo{author}{\bibfnamefont{M.}~\bibnamefont{Russina}},
  \bibinfo{author}{\bibfnamefont{A.}~\bibnamefont{Furrer}},
  \bibinfo{author}{\bibfnamefont{A.}~\bibnamefont{Alfonsov}},
  \bibinfo{author}{\bibfnamefont{E.}~\bibnamefont{Vavilova}},
  \bibinfo{author}{\bibfnamefont{V.}~\bibnamefont{Kataev}},
  \bibinfo{author}{\bibfnamefont{B.}~\bibnamefont{B{\"u}chner}},
  \bibinfo{author}{\bibfnamefont{T.}~\bibnamefont{Str{\"a}ssle}},
  \bibinfo{author}{\bibfnamefont{E.}~\bibnamefont{Pomjakushina}},
  \bibinfo{author}{\bibfnamefont{K.}~\bibnamefont{Conder}},
  \bibnamefont{et~al.}, \bibinfo{journal}{Phys. Rev. Lett.}
  \textbf{\bibinfo{volume}{101}}, \bibinfo{pages}{247603}
  (\bibinfo{year}{2008}).

\bibitem[{\citenamefont{Podlesnyak et~al.}(2011)\citenamefont{Podlesnyak,
  Ehlers, Frontzek, Sefat, Furrer, Str{\"a}ssle, Pomjakushina, Conder, Demmel,
  and Khomskii}}]{2011_Podlesnyak}
\bibinfo{author}{\bibfnamefont{A.}~\bibnamefont{Podlesnyak}},
  \bibinfo{author}{\bibfnamefont{G.}~\bibnamefont{Ehlers}},
  \bibinfo{author}{\bibfnamefont{M.}~\bibnamefont{Frontzek}},
  \bibinfo{author}{\bibfnamefont{A.~S.} \bibnamefont{Sefat}},
  \bibinfo{author}{\bibfnamefont{A.}~\bibnamefont{Furrer}},
  \bibinfo{author}{\bibfnamefont{T.}~\bibnamefont{Str{\"a}ssle}},
  \bibinfo{author}{\bibfnamefont{E.}~\bibnamefont{Pomjakushina}},
  \bibinfo{author}{\bibfnamefont{K.}~\bibnamefont{Conder}},
  \bibinfo{author}{\bibfnamefont{F.}~\bibnamefont{Demmel}}, \bibnamefont{and}
  \bibinfo{author}{\bibfnamefont{D.~I.} \bibnamefont{Khomskii}},
  \bibinfo{journal}{Phys. Rev. B} \textbf{\bibinfo{volume}{83}},
  \bibinfo{pages}{134430} (\bibinfo{year}{2011}).

\end{thebibliography}

%

\end{document}